\documentclass[letterpaper, 10 pt, conference]{ieeeconf}  

\IEEEoverridecommandlockouts                              

\usepackage{amsfonts}
\usepackage{amsmath}
\usepackage{mathtools}
\usepackage{mathrsfs}
\usepackage{xcolor}
\usepackage[linesnumbered,ruled,vlined]{algorithm2e}
\usepackage{graphicx}
\usepackage{relsize}
\usepackage{amssymb}

\usepackage{algpseudocode}
\newtheorem{remark}{Remark}

\SetCommentSty{mycommfont}
\SetKwInput{KwInput}{Input}                
\SetKwInput{KwOutput}{Output}   
\title{\LARGE \bf
An Iterative Approach to Data-Driven Inference for Decoding Oscillator Network Structures
}

\author{Shicheng Li$^{a,\dagger}$, Bharat Singhal$^{b,\dagger}$,  and Jr-Shin Li$^{b,c}$
\thanks{$^\dagger$ Both authors contribute equally and share the first authorship.}
\thanks{$^a$ Department of Mechanical Engineering and Materials Science, Washington University in St. Louis, St. Louis, MO, USA. } 
\thanks{$^b$ Department of Electrical and Systems Engineering, Washington University in St. Louis, St. Louis, MO, USA. }
\thanks{$^c$ Division of Biology and Biomedical Sciences, Washington University in St. Louis, St. Louis, MO, 63130, USA. Correspondence to: jsli@wustl.edu.} \\
\thanks{$^*$This work was supported by 
the Air Force Office of Scientific Research under the award FA9550-21-1-0335.}
}

\begin{document}

\setlength{\abovedisplayskip}{5pt}
\setlength{\belowdisplayskip}{5pt}
\maketitle
\thispagestyle{empty}
\pagestyle{empty}

\begin{abstract}                
In complex networks, interactions between multiple agents give rise to an array of intricate global dynamics, ranging from synchronization to cluster formations. Decoding the connectivity structure as well as the types of interactions from measurement data is the first step toward understanding these intriguing behaviors. In this paper, we present a bilinear optimization framework to infer both the connectivity and interaction functions of oscillator networks with the identical class of coupling functions. We then propose an iterative algorithm to solve the resulting bilinear problem and illustrate its convergence. We validate our approach on both simulated and noisy experimental datasets, where we demonstrate its effectiveness compared with existing approaches.  
\end{abstract}


\section{Introduction}
\label{sec:Intro}
Complex systems with multiple interacting agents are prevalent in nature and human society on different scales~\cite{strogatz2001exploring}. The dynamic interplay exhibited by these interconnected agents gives rise to a multitude of intricate network functions, the understanding of which requires decoding the connectivity and the specific nature of interactions, i.e., coupling functions, between the agents. For example, elucidating the neuronal network in the suprachiasmatic nucleus (SCN) is critical for our understanding of the mechanisms driving SCN synchronization~\cite{abel2016functional}, while determining the changes in the brain network structure is important to predict the onset of seizures~\cite{bomela2020real}. In these scenarios, time series recording from each agent (node) is accessible, and the objective is to determine both the network structure and the nature of interactions. 

The aforementioned applications have led to the development of several network inference techniques. The fundamental idea behind these methods is to learn the network model, i.e., the network topology and coupling functions, that explain the recorded time series data\footnote{It is important to note that the information-theoretic measures such as correlation, or mutual-information, are not considered as they determine the functional connectivity, not the effective connectivity (see \cite{friston2011functional} for the difference between functional and effective connectivity) }. The primary difference arises from the variations in the candidate network models and the approach used to identify the best-fit model. For example, inference methods employing Granger causality typically fit a vector autoregressive model to the time series, and the autoregression coefficients are determined using classical likelihood ratio tests~\cite{friston2011functional,lu2021causal}. Similarly, regression-based methods transform the network inference problem into a least-squares (or LASSO~\cite{tibshirani1996regression}) problem by utilizing the orthonormal function approximation~\cite{wang2018inferring,shandilya2011inferring,Singhal2023}.

In numerous experimental and physiological scenarios, nodes within the network are interconnected through similar forms of interaction (coupling) functions.  For example, electrochemical oscillator networks have resistive coupling~\cite{kiss2002emerging}, and neurons in the SCN communicate primarily through VIP and GABA signaling~\cite{aton2005vasoactive}. These applications necessitate inference techniques that can utilize the relationship between the network coupling functions of different nodes. Indeed, the inability to harness these relationships would result in suboptimal performance of inference techniques.


In this paper, employing this observation, we consider oscillator networks with similar coupling functions and propose a bilinear optimization formulation for the network inference problem. Specifically, we first transform the network inference task into a bilinear optimization by leveraging the Fourier decomposition of orthogonal functions. We then introduce an iterative algorithm to solve the resulting bilinear problem and substantiate its convergence properties. We compare our method to some of the existing network inference techniques and show its effectiveness on networks of Kuramoto and Stuart-Landau oscillators. We further validate our technique on experimental data obtained from a network of nonlinear electronic oscillators, illustrating its robustness against measurement noise. 
\section{Background and Problem Formulation}
In this section, we introduce the network model for coupled limit-cycle oscillators and formulate the network inference problem. 
\subsection{Background}
We consider a network of $N$ limit-cycle oscillators such that the dynamics of node $i$ comprises its own dynamics and the pair-wise coupling with other nodes in the network, i.e.,
\begin{equation}
    \label{eq:network_model}
	\frac{d}{dt}X_i(t) = f_i(X_i(t)) + \sum_{\substack{j=1 \\j \neq i}}^N \epsilon_{ij}K(X_i(t),X_j(t)).
\end{equation}
Here, $X_i\in \mathbb{R}^n$ $(n\geq2)$, $f_i$ are the state and self-dynamics of oscillator $i$, $K$ is the coupling function, and $\epsilon_{ij}$ scales the strength of connection from node $j$ to $i$. If two nodes are uncoupled then $\epsilon_{ij}=\epsilon_{ji}=0$. For the network \eqref{eq:network_model}, if the coupling strengths are sufficiently weak then a reduced one-dimension model capturing the phase dynamics of each oscillator can be obtained by leveraging the phase-reduction theory (see ref. \cite{schwemmer2012theory} for more details). As a result, the phase dynamics of oscillator $i$ can be written as 
\begin{equation}
    \label{eq:Network_phase_model}
    \dot{\varphi_i}(t) = \omega_i+\sum_{\substack{j=1\\ j\neq i}}^N k_{ij}\alpha(\varphi_j(t)-\varphi_i(t))
\end{equation}
where $\varphi_i, \omega_i \in \mathbb{R}$ are the phase and natural frequency of the oscillator $i$, $\alpha$ is a $2\pi-$ periodic coupling function, and $k_{ij}$ denote the strength of connection from node $j$ to node $i$, if there is a connection, otherwise $k_{ij}=0$. It is important to note that our objective is not to infer the coupling functions $K$ of network \eqref{eq:network_model}. Instead, we aim to recover the phase coupling function $\alpha$ of \eqref{eq:Network_phase_model} as accurate inference of both the coupling function and the network connectivity (represented by $\{k_{ij}\}$) is crucial to predict the synchronization behavior and to control the network functionality, for instance, network desynchronization.

\subsection{Network Inference Problem}
Given the time series recordings from each node $\{x_i(t_k)\}$, where $i=1,\dots,N$ and $k=1,\dots,M+1$, and $x_i$ is the one-dimensional observable of oscillator state $X_i$, we want to efficiently (with limited measurements) infer the connectivity structure $\{k_{ij}\}$ and phase coupling functions of the network.

\section{Theory and Algorithm} 
In this section, we present the bilinear optimization formulation of the network inference problem and the iterative algorithm to solve the resulting optimization problem. 

To begin, we first obtain the phase trajectories of each oscillator from their time series data. These phases can be determined using conventional phase estimation algorithms such as peak-finding or wavelet transform~\cite{mitrou2017peak}. The next step is to infer the unknown parameters $\{k_{ij}\}$ and the function $\alpha$ of phase model \eqref{eq:Network_phase_model} by plugging in the estimated phases. To this end, without loss of any generality, we approximate the $2\pi-$ periodic smooth coupling function using truncated Fourier series, i.e., 
\begin{equation}
    \label{eq:coupling_function_approximation}
     \alpha(\Delta\varphi_{ij}) \approx a_{0} + \sum_{m=1}^{r} a_{m}\cos(m(\Delta\varphi_{ij})) + b_{m}\sin(m(\Delta\varphi_{ij})) ,
\end{equation}
where $\Delta\varphi_{ij} = \varphi_j-\varphi_i$, $r$ is the number of Fourier harmonics that can be taken sufficiently large to approximate the coupling function with arbitrary accuracy, and $\{a_m,b_m\}$ are the unknown Fourier coefficients. As a result of this Fourier decomposition, the phase dynamics \eqref{eq:coupling_function_approximation} can be expressed as 
\begin{equation}
    \label{eq:phase_dynamics_fourier}
    \dot{\varphi_i}(t) \approx \bar{\omega}_i+\sum^N_{j=1\neq i}k_{ij}\sum_{m=1}^r a_{m} \cos(m\Delta \varphi_{ij})+b_{m} \sin(m\Delta \varphi_{ij}),
\end{equation}
where $\bar{\omega}_i = \omega_i+\sum^N_{j=1, j\neq i}k_{ij}a_{0}$ is the adjusted frequency. On a closer inspection, we observe that \eqref{eq:phase_dynamics_fourier} is in a bilinear matrix form 
\begin{equation}
    \label{eq:bilinear_matrix_form}
    \dot{\varphi_i}(t) = K_i^TZ_i(t)A+\epsilon_i(t),
\end{equation}
where $\epsilon_i(t)$ corresponds to the approximation error, $K = [k_{i1},\dots,k_{iN},\bar{\omega}_i]^T \in \mathbb{R}^N$ contains the incoming connections to node $i$, $A=[a_1, \dots, a_r , b_1 , \dots , b_r , 1]^T \in  \mathbb{R}^{2r+1}$ represents the coupling function coefficients, and 
\begin{equation*}
    \label{eq:Z_matrix}
    Z_i(t) = \left[
    \begin{smallmatrix}
        \cos(\Delta \varphi_{i1}) & \dots & \cos(r\Delta \varphi_{i1}) & \sin(\Delta \varphi_{i1}) & \dots & \sin(r\Delta \varphi_{i1}) & 0 \\
        \vdots &  & \vdots & \vdots &  & \vdots  & \vdots \\ 
        \cos(\Delta \varphi_{iN}) & \dots & \cos(r\Delta \varphi_{iN}) & \sin(\Delta \varphi_{iN})& \dots & \sin(r\Delta \varphi_{iN}) & 0 \\
        0 & \dots & 0 & 0 & \dots & 0 & 1
    \end{smallmatrix} \right].   
\end{equation*}

By combining \eqref{eq:bilinear_matrix_form} for all sample points $t_1,\dots,t_{M+1}$, we obtain a set of $M$ bilinear equations, where $\dot{\varphi_i}$ is obtained by the first-order approximation of the derivative, i.e., $\dot{\varphi_i}(t_k) = \frac{\varphi(t_{k+1})-\varphi(t_k)}{t_{k+1}-t_k}$, $k=1,\dots,M$. The network inference problem (for node $i$) can now be formulated as an optimization problem where the estimates of the unknown coefficient vectors $K_i$ and $A$ are determined by  
\begin{equation}
    \label{eq:bilinear_minimization_problem}
  (\hat{K_i},\hat{A}) =  \arg \min\limits_{K_i,A} \mathcal{L}_i(K_i,A), 
\end{equation}
where $\mathcal{L}_i(K_i,A) = \sum_{k=1}^M (\dot{\varphi_i}(t_k) - K_i^TZ_i(t_k)A )^2$.

We want to emphasize that our network inference formulation is different from the existing works using orthonormal functional approximation~\cite{wang2018inferring,shandilya2011inferring} as we assume the coupling functions $\alpha$ to be identical and vary the strengths of the connections. Specifically, existing works model the connection between node $i$ and $j$ as $\alpha_{ij}(\Delta\varphi_{ij})$, different from our formulation where we take it as $k_{ij}\alpha(\Delta\varphi_{ij})$ with $k_{ij}$ being an unknown constant.  As a result, the network inference problem turns into a bilinear least-squares form, different from the linear least-squares in existing literature. The rationale behind this model consideration, as explained in section \ref{sec:Intro}, is to facilitate the efficient recovery of network parameters by reducing the number of unknown parameters from $N^2\times2r$ to $N^2+2r$. 


\begin{remark}
Note that if $(\hat{K}_i,\hat{A})$ is a solution of \eqref{eq:bilinear_minimization_problem}, so is $(c\hat{K}_i,\hat{A}/c)$ for any $c\neq0$. To avoid this ambiguity, we normalize $A$, i.e., $\Vert A \Vert_2=1$, and assume the $\hat{K}_i$ to have non-negative elements as the coupling strength cannot be negative. 
\end{remark}

\subsection{Iterative Algorithm for Network Inference}
In this section, we propose an iterative method to solve the bilinear optimization problem \eqref{eq:bilinear_minimization_problem} while following the non-negative $\hat{K}_i$ constraint. The principle idea of our algorithm is to solve \eqref{eq:bilinear_minimization_problem} for each node iteratively. In the following, we describe the network inference algorithm in detail. 

 We start by arbitrarily initializing Fourier coefficients of node 1, $\hat{A}_1^{(0)}$ such that $\Vert \hat{A}_1^{(0)}\Vert_2 = 1$ and then obtain connectivity coefficients $\hat{K}^{(1)}_1$ after plugging in $\hat{A}_1^{(0)}$ in \eqref{eq:bilinear_minimization_problem}. Specifically,
\begin{equation*}
    \hat{K}^{(1)}_1 = \arg \min\limits_{K_1\geq 0}\mathcal{L}_1(K_1,\hat{A}_1^{(0)}) = \arg \min\limits_{K_1\geq 0} \Vert Y_1-X_{1,A}^{(0)}K_1\Vert_2^2,
\end{equation*}
where 
\begin{equation*}
    Y_1=\begin{bmatrix}
        \dot{\varphi}_1(t_1) \\
        \vdots \\
        \dot{\varphi}_1(t_M)
    \end{bmatrix} , \qquad X_{1,A}^{(0)} = \begin{bmatrix}
        (Z_1(t_1)A_1^{(0)})^T \\ 
        \vdots \\
        (Z_1(t_M)A_1^{(0)})^T
    \end{bmatrix},
\end{equation*}
which is a non-negative least-squares problem. Now the estimated $\hat{K}^{(1)}_1$ are used to update the Fourier coefficients, i.e., $\hat{A}_1^{(1)} = \arg \min\limits_{A} \mathcal{L}_1(K_1^{(1)},A)$ (the superscript denotes the iteration number). The updated Fourier coefficients can be directly obtained as 
\begin{equation*}
    \hat{A}_1^{(1)} = ({X_{1,K}^{(1)}}^TX_{1,K}^{(1)})^{-1} {X_{1,K}^{(1)}}^T Y_1,
\end{equation*}
where 
\begin{equation*}
     X_{1,K}^{(1)} = \begin{bmatrix}
        {K_1^{(1)}}^TZ_1(t_1)\\ 
        \vdots \\
        {K_1^{(1)}}^TZ_1(t_M)
    \end{bmatrix}.
\end{equation*}
This process is repeated till the Fourier coefficient converges, i.e., 
$\hat{A}_1^{(I_1)}-\hat{A}_1^{(I_1-1)}<\epsilon,$ where $\epsilon$ is the user-defined threshold and $I_1$ is the number of iterations. The output Fourier coefficients at $I_1$ iteration are then used as an initialization to solve \eqref{eq:bilinear_minimization_problem} for node 2. The principle idea behind this sequential initialization is to leverage the fact that all nodes have identical Fourier coefficients. This process of sequential initialization, i.e., $\hat{A}_i^{(I_i)} = \hat{A}_{i+1}^{(0)}, i=1,\dots,N-1$ is executed for all nodes.  
\begin{algorithm}[!ht]
    \DontPrintSemicolon
    \KwInput{Normalized $A_1^{(0)}$, $\epsilon$, and $\Lambda$}
    \KwOutput{$\hat{A}$, $\hat{K}_i$ for $i=1,\dots,N$}
    \KwData{$\varphi_i(t_k)$ and $\dot{\varphi}_i(t_k)$ for $k=1,\dots,M$}
    $\lambda=1$ \tcp*{$\lambda$: network iteration number }
    \While{$\lambda<\Lambda$}
    {
    $i,j = 1$   \tcp*{$i,j$: node and iteration number}
    $\hat{K}^{(j)}_i = \arg \min\limits_{K_i\geq 0}\mathcal{L}_i(K_i,\hat{A}_i^{(j-1)})$\;
    $\hat{A}^{(j)}_i = \arg \min\limits_{A}\mathcal{L}_i(\hat{K}^{(j)}_i,A)$ \;
    Normalize $\hat{A}^{(j)}_i$\;
    \If{$\Vert\hat{A}^{(j)}_i-\hat{A}^{(j-1)}_i  \Vert_2 \leq \epsilon$}
        {
        \If{$i==N$} 
            {
            \tcp*{Solve \eqref{eq:bilinear_minimization_problem} for node 1.}
            $\lambda=\lambda+1$ and $I_N=j$\;
            $\hat{A}_1^{(0)}=\hat{A}_N^{(j)}$  and go to step 2. 
            }
        \Else 
            {
            \tcp*{Solve \eqref{eq:bilinear_minimization_problem} for the next node.}
            $I_i=j$, $i=i+1$, and $j=1$\; 
            $\hat{A}_i^{(0)}=\hat{A}_{i-1}^{(I_{i-1})}$ and go to step 4.
            }
        }
    \Else
        {
        $j=j+1$ and go to step 4.
        }
    } 
    Estimate $\hat{A} = \frac{1}{N}\sum_{i=1}^{N}A_i^{(I_i)}$ \;
    Estimate $\hat{K}_i = \arg \min\limits_{K_i\geq 0}\mathcal{L}_i(K_i,\hat{A}) \quad i=1,\dots,N.$
    \caption{Estimation of $A$ and $K_i$, $i=1,\dots,N$.} 
\label{algorithm-1} 
\end{algorithm}

 Note that since we start from the random initialization, i.e., $\hat{A}_1^{(0)}$, the Fourier coefficients for the first few nodes might deviate from the true Fourier coefficients. To resolve this initial random initialization issue, the converged Fourier coefficients for the last node, $\hat{A}_N^{(I_N)}$, are then assigned as the initialization of the first node, $\hat{A}_1^{(0)}$, and the whole process is repeated for all the nodes again. This looping process, i.e., $\hat{A}_N^{(I_N)}=\hat{A}_1^{(0)}$ is performed till we have performed a prescribed number of network iterations $\Lambda$ (see Algorithm~\ref{algorithm-1}).    

Finally, to ensure all nodes have identical Fourier coefficients, we take 
\begin{equation}
    \hat{A} = \frac{1}{N}\sum_{j=1}^N \hat{A}_i^{(I_i)}.
\end{equation}
This averaging also removes the estimation errors in different $\hat{A}_i^{(I_i)}$. Now that $A$ has been estimated, we determine $\hat{K}_i, i=1,\dots,N$, by solving the non-negative least-squares problem 
\begin{equation*}
    \hat{K}_i = \arg \min\limits_{K_i\geq 0}\mathcal{L}_i(K_i,\hat{A}).
\end{equation*}
This whole procedure is outlined in Algorithm~\ref{algorithm-1}.

 \begin{remark}
 It has been shown that the iterative procedure to solve \eqref{eq:bilinear_minimization_problem} converges to a stationary point (see \cite{bai2004convergence}), given sufficiently large data points. Moreover, under particular conditions, the algorithm converges in only one step. However, different initialization might yield a different stationary point. The proposed sequential initialization resolves this bottleneck, as observed in our numerical studies.  
 \end{remark}

\subsection{Convergence Analysis}
In this section, we present a numerical analysis of the convergence of the proposed algorithm. To keep the analysis informative, we consider a network of $5$ Kuramoto oscillators, i.e., $N=5$, with oscillator $i$ following the phase dynamics 
\begin{equation}
    \label{eq:Kuramoto_oscillators}
    \dot{\theta}_i = \omega_i + \sum_{j=1}^N k_{ij} \sin{(\theta_j-\theta_i)},
\end{equation}
where $\theta_i$ and $\omega_i$ are the phase and nature frequency; $k_{ij}$ is the strength of the sinusoidally coupled connection with node $j$. The natural frequencies of the oscillators are uniformly distributed in $[3,5]$ rad/s and $k_{ij} \in [0.1,0.3]$, if node $i$ and $j$ are connected. We consider $10$ such random networks and for each network, time series data for $5$ cycles, i.e., $(T=2\pi\/)$ with a total of $200$ sampling points is generated. We then apply our inference algorithm for each network, where the coupling function is approximated using two Fourier harmonics, i.e., $r=2$. The value of the algorithm parameters $\epsilon$ and $\Lambda$ are set as $10^{-5}$ and $2$ and the Fourier coefficients are initialized to $\hat{A}_1^{(0)} = (0.25,0.25,0.25,0.25)^\prime$. The error in the estimated Fourier coefficients for each node $\Vert\hat{A}_i^{(I_i)}-A\Vert_2$ is plotted in Figure 1, where the true Fourier coefficients $A=(0,0,1,0)^\prime$. The blue (black) line corresponds to the error in the Fourier coefficients when $\lambda=1$ $(\lambda=2)$. We find that in the second round of network iteration ($\lambda=2$), the error drops to zero for all $10$ networks, i.e., the Fourier coefficients converge to their true value  The average number of iterations for each node remains low. Specifically, For the first network iteration $\{I_1,I_2,I_3,I_4,I_5\} = \{75,42,26,32,17 \}$, and for the second network iteration $\{13,3,12,6,3 \}$. 
\begin{figure}[hbt!]
	\centering
	\includegraphics[scale=0.66]{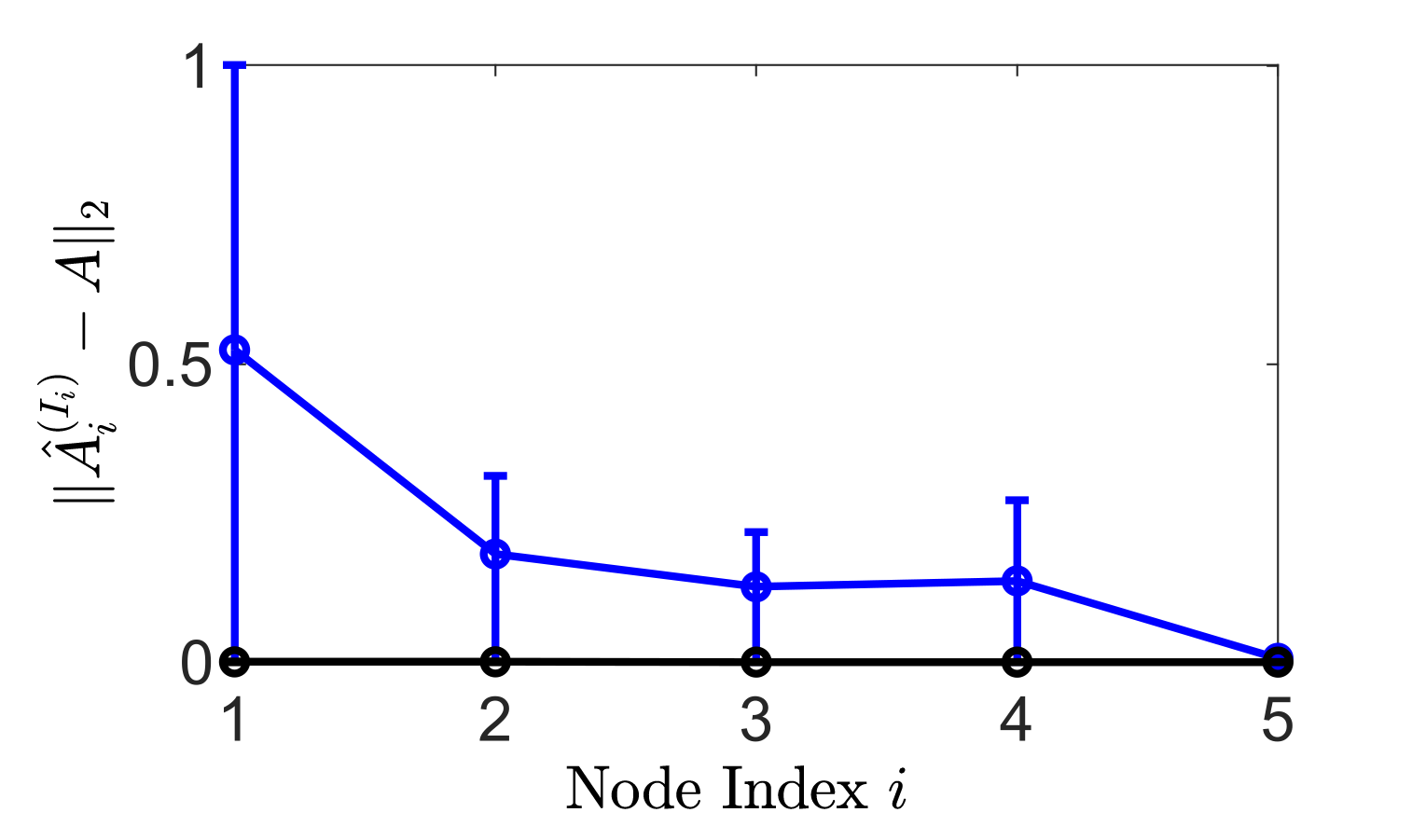}
	\caption{Numerical convergence analysis of the proposed algorithm. The blue (black) line denotes the mean error in the Fourier coefficients in the first (second) network iteration. Error bars represent the first and third quartiles. }.
	\label{fig:figure_1}
\end{figure}

The presented numerical analysis highlights the importance of our sequential initialization scheme, which results in the convergence of $\hat{A}$ to its true value. In addition, the number of iterations per node decreases as $\lambda$ increases. 

\section{Numerical Simulations and Analysis}
In this section, we illustrate the ability of our algorithm to infer large oscillator networks efficiently compared to the commonly used inference techniques. To this end, we consider the networks of Kuramoto and Stuart-Landu oscillators, which are widely used to model the oscillatory behavior of complex systems. For the comparison purpose, we use ICON~\cite{wang2018inferring} and sparse regression which involves applying $l_1$ penalty on the unknown coefficients and has been shown effective for the inference of dynamic topology~\cite{timme2014revealing,napoletani2008reconstructing}. To begin, we describe the performance evaluation metrics that we use to analyze the accuracy of inferred networks.

\subsection{Performance Evaluation Metric}
We use two different criteria for the comparative analysis of our technique: (i) estimation error in the coupling function measured by $2\int_0^{2\pi} \left \Vert \alpha(\Delta\theta)-\hat{\alpha}(\Delta\theta) \right\Vert_2^2 d\Delta\theta = \Vert A-\hat{A}\Vert_2^2$, and (ii) ability of the algorithm in recognizing true and false connections measured by area under the receiver operating characteristic curve (ROC curve). The ROC curve plots the variation of the true positive rate (TPR) with the false positive rate (FPR) as the classification threshold is varied~\cite{Singhal2023}. A greater area under the ROC curve (AUC score) means that the inference algorithm is better at classifying true and false connections. The quantities TPR and FPR measure the proportion of correctly identified links and the proportion of links that were missed by the inference algorithm (see ref. \cite{hanley1982meaning} for more details). 

\subsection{Kuramoto Oscillator Network}
We consider a network of 100 Kuramoto oscillators $(N=100)$ where the dynamics of each node are described by $\eqref{eq:Kuramoto_oscillators}$. The natural frequencies of the oscillators are uniformly distributed in $[3,5]$ rad/s, and the coupling strengths $k_{ij}\in [0.05,0.1]$. We generate $10$ such networks with Erd{\H o}s-R{\'e}nyi ~\cite{renyi1960evolution} topology, and for each network, we simulate the network dynamics for $5$ times. In each simulation, the network starts from random initial conditions, and the data is collected for $10$ cycles, i.e., $T=4\pi$, with a total of 1000 sampled data points. 

\begin{figure}[hbt!]
	\centering
	\includegraphics[scale=0.6]{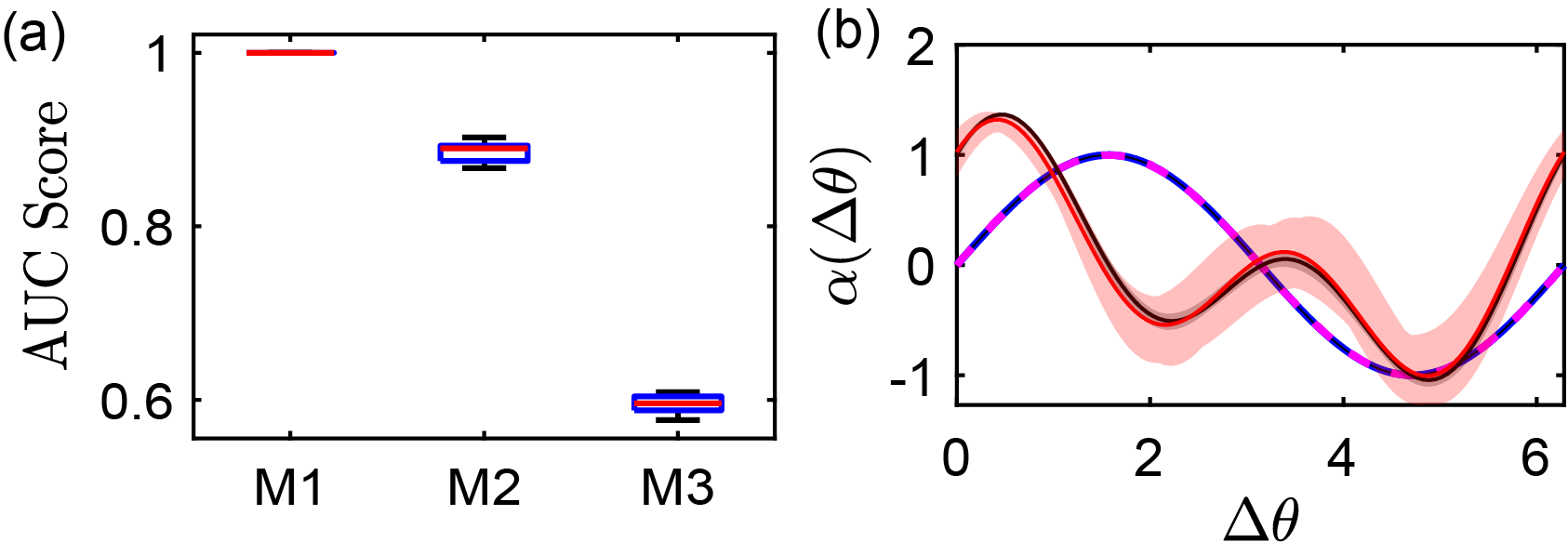}
	\caption{Comparison of the proposed technique with other inference methods on a network of Kuramoto oscillators. (a) Box plot illustrating the estimated AUC scores for 10 randomly generated networks of 100 nodes. M1: Proposed technique, M2: sparse regression, and M3: ICON. (b) Mean estimated coupling function of the candidate networks using M1 (blue line), M2 (black), and M3 (red), where the dotted magenta line denotes the true coupling function. The shaded region denotes the variation in the estimated coupling functions across different networks.  }.
	\label{fig:figure_2}
\end{figure}

Now, we apply the proposed technique along with ICON and sparse regression. The error threshold $\epsilon$ is set as $10^{-5}$ and the vector $\hat{A}_1^{0}$ is initialized as a unit vector with identical coefficients. The number of Fourier approximation terms $r$ is kept as 2 for all three methods. Figure~\ref{fig:figure_2}(a) shows the box plots of the estimated AUC scores for all 10 networks. The proposed method gets an AUC score of one for all the networks. The sparse regression gets the second-best AUC score with a mean of 0.88; while ICON recovered networks have a mean AUC score of 0.60. In addition to accurately determining the network connectivity structure, our technique predicts the coupling function precisely; while the sparse regression and ICON coupling function do not reflect the true coupling dynamics (Figure~\ref{fig:figure_2}(b)). The errors in the coupling function estimation $\Vert A-\hat{A} \Vert_2^2 $ are in  $[1.8\times10^{-21},1.3,\times10^{-11}]$, $[0.9,1.1]$, and $[0.8,1.6]$ for the proposed method, sparse regression, and ICON, respectively. Note that we take the average of all $\alpha_{ij}(\Delta\theta)$, $i,j=1,\dots,N$, for the sparse regression and ICON as the individual coupling function between the nodes can be different. 

\begin{figure}[hbt!]
	\centering
	\includegraphics[scale=0.7]{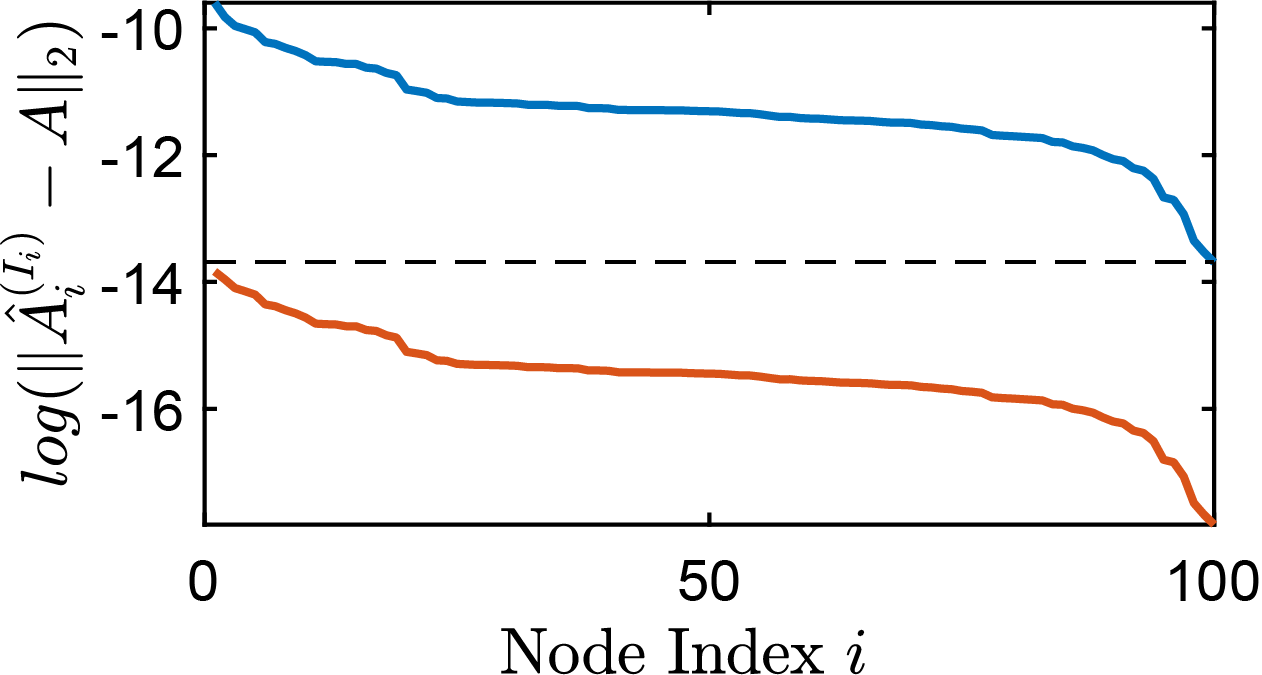}
	\caption{Convergence behavior of the proposed algorithm for a 100-node Kuramoto network. The blue (red) line denotes the error in the estimated Fourier coefficients in the first (second) network iteration.  }.
	\label{fig:figure_3}
\end{figure}

The convergence results of the estimated Fourier coefficients to the true Fourier coefficients for a randomly selected network out of the 10 candidate networks are shown in Figure~\ref{fig:figure_3}. The blue (red) line denotes the error in the Fourier coefficients for the first (second) network iteration. The average number of iterations per node also remains low ($\approx 3$). 

\begin{figure}[hbt!]
	\centering
	\includegraphics[scale=0.65]{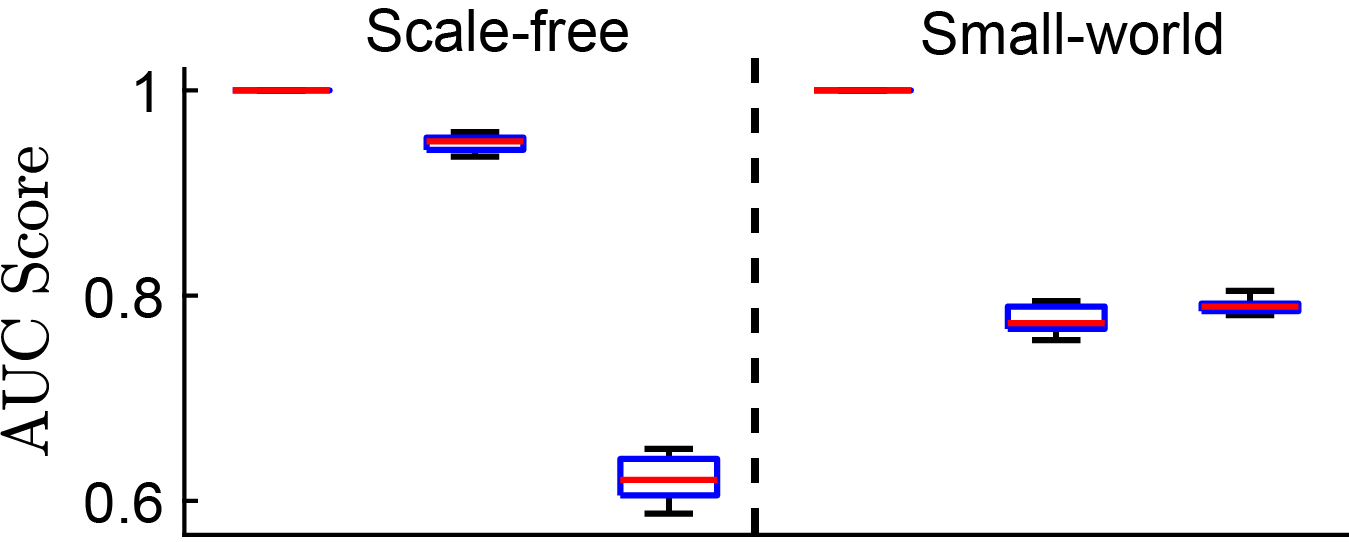}
	\caption{Effect of network topology on the accuracy of network reconstruction. The first, second, and third barplot of each section correspond to the proposed method, sparse regression, and ICON, respectively. }
	\label{fig:figure_topology}
\end{figure}

We also examine the effect of network topology on inference accuracy by considering networks of Small-world~\cite{watts1998collective} and BA Scale-free~\cite{albert2002statistical} topologies, in addition to the  Erd{\H o}s-R{\'e}nyi topology of Figure~\ref{fig:figure_1}. We generate 10 networks (100 nodes) for each topology with similar model parameters to Figure~\ref{fig:figure_1} and do the comparative analysis. We find that our algorithm performs consistently for both network topologies; while the inference accuracy of sparse regression degrades for small-world topology (Figure~\ref{fig:figure_topology}). This demonstrates the robustness of our approach to the underlying network structure. 

\subsection{Stuart Landau (SL) Oscillator Network}
So far, we demonstrated the performance of our technique on a coupled phase model (Kuramoto network). Now, we consider the SL oscillator model where the phases are not directly available but are estimated from the measured time series data. The SL oscillator is widely employed to model complex oscillatory behavior in a variety of applications, ranging from semiconductor lasers to pacemaker cells in the heart~\cite{wang2021collective}. A network of $N$ SL oscillators is described by 
\begin{align*}
    & \dot{x}_i(t) = ax_i-\omega y_i - (x_i^2+y_i^2)x_i + \epsilon_{ij} \sum_{j=1}^{N} (x_j-x_i), \\
    & \dot{y}_i(t) = \omega x_i+ay_i - (x_i^2+y_i^2)y_i + \epsilon_{ij} \sum_{j=1}^{N} (y_j-y_i), 
\end{align*}
where $x_i(t)$ and $y_i(t)$ are the states of oscillator $i$, and $\epsilon_{ij}\in [0.05,0.1]$ denotes the strength of the connection between node $i$ and $j$ if it exists, otherwise $\epsilon_{ij}=0$. The model parameters $a$ and $\omega$ determine the amplitude and frequency of oscillation, respectively. We consider $10$ such networks of $100$ nodes ($a=1$, $\omega=1$), and for each network, we generate data by stimulating the model $50$ times from random initial conditions. In each simulation, we collect the data for $7$ cycles with a sampling rate of $200$ points per cycle. Note that the data is generated from multiple simulations to avoid network synchronization~\cite{shandilya2011inferring}. This is because inference of topology becomes infeasible when the network is synchronized since the observed phase difference becomes constant and no longer provides any additional information ($Z_i(t)$ will not change with time during synchronization).
\begin{figure}[hbt!]
	\centering
	\includegraphics[scale=0.6]{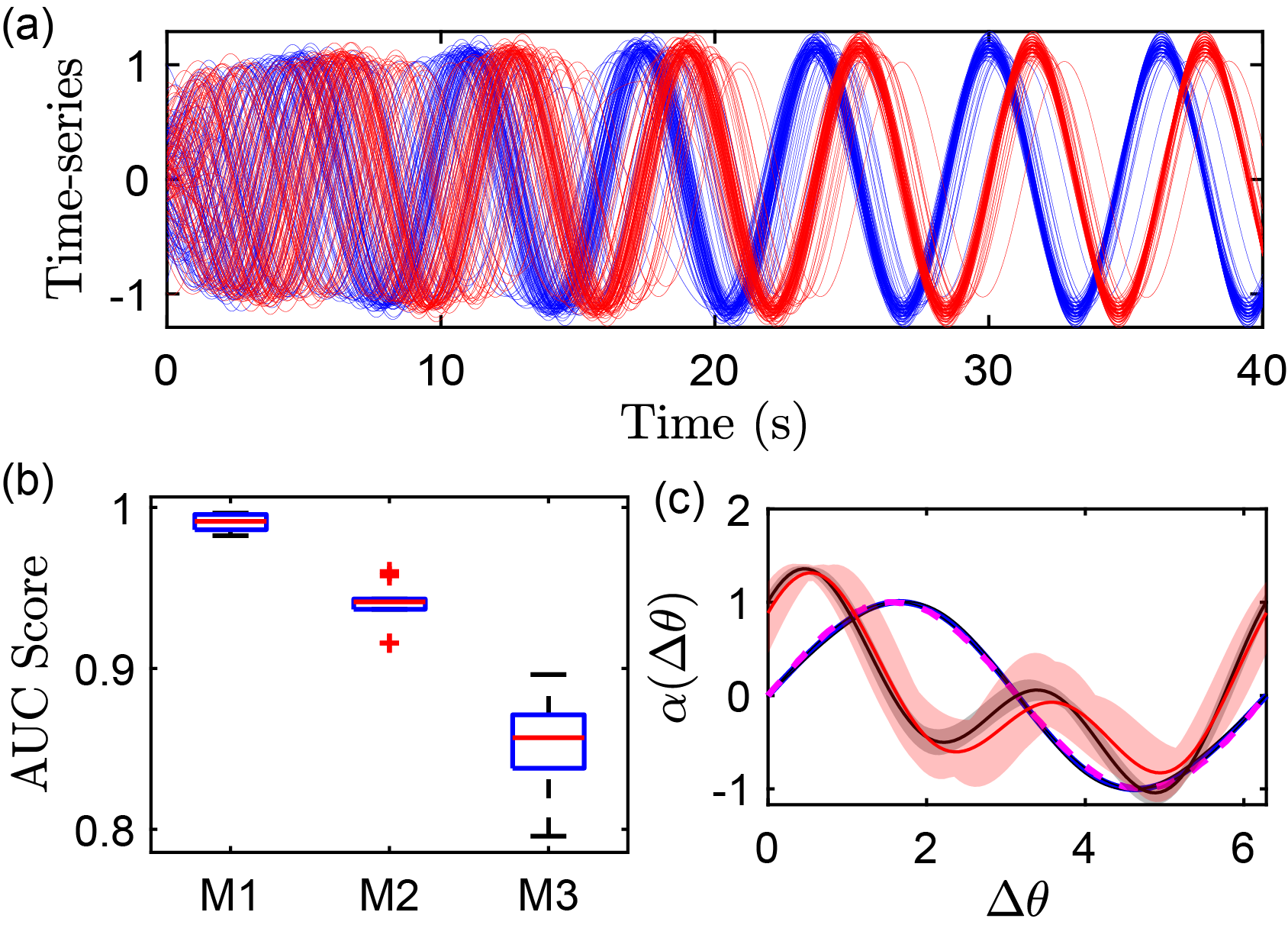}
	\caption{Comparative analysis of the proposed technique with ICON and sparse regression on a network of SL oscillators. (a) Time series data displaying the synchronization behavior of the network. The red and blue lines denote the states $x_i(t)$ and $y_i(t)$ for $i=1,\dots,100$. (b) Box plot illustrating the estimated AUC scores for 10 randomly generated networks of 100 nodes. M1: Proposed technique, M2: sparse regression, and M3: ICON. (c) Mean of the estimated coupling functions using M1 (blue line), M2 (black), and M3 (red), where the dotted magenta line denotes the true coupling function. The shaded region denotes the variation in the coupling functions across different networks.   }.
	\label{fig:figure_4}
\end{figure}

After generating the data for each network, we estimate the phase from the states $x_i(t)$ and $y_i(t)$ by $\varphi_i(t) = \arctan (y_i(t) / x_i(t))$~\cite{nakao2016phase} and apply the proposed inference method ($\epsilon=10^{-5},\Lambda=2,r=2, \text{ and } \hat{A}_1^{0}$ is initialized as a unit vector with identical elements) along with ICON and sparse regression. The results are shown in Figure~\ref{fig:figure_4}, where the panel b displays the bar plots of the estimated AUC scores. The proposed method outperforms both sparse regression and ICON; the mean AUC scores for our algorithm, sparse regression, and ICON are 0.99, 0.93, and 0.85, respectively. The errors in the coupling function estimation are also significantly lower for our method. Specifically, $\Vert A-\hat{A}\Vert_2^2\in[6\times 10^{-4},7\times10^{-3}]$ for our technique compared with $[0.8,1.2]$ for sparse regression and $[0.6,1.6]$ for ICON. The true phase coupling function of the SL oscillator was estimated using phase-reduction theory~\cite{nakao2016phase}.   

\section{Experimental Validation}
In this section, we consider experimental data obtained from a network of 28 R\"{o}ssler electronic oscillators and evaluate the robustness of our technique to measurement noise - an uncontrollable element in an experimental setting. This experimental dataset contains time series recordings from 20 distinct network topologies. Notably, for each network topology, data was recorded under various coupling strengths between the oscillators, with three separate time series acquired for each combination of coupling strength and network configuration~\cite{vera2020experimental}. 

We first obtain the phases of the oscillators from the measured data using the peak-finding method~\cite{mitrou2017peak}. The principle idea behind the peak-finding technique is to assign a phase $2\pi k$ at time $t_k$, the $k^{th}$ peak of the time series data, and then estimate the phase at time $t$, ($t_{k}<t<t_{k+1}$) using linear interpolation. 
After obtaining the phase data, we apply Algorithm~\ref{algorithm-1} ($\epsilon=10^{-5}$, $\Lambda=2$, $r=1$, and $\hat{A}_1^{(0)}$ is a unit vector of identical elements) along with ICON and sparse regression to all $20$ network topologies for two coupling strengths $k_{ij}=0.01$ and $k_{ij}=0.1$. The estimated AUC scores are all these network combinations are shown in Figure 4, where the proposed technique outperforms both ICON and sparse regression for all the networks even with the noisy data. 
\begin{figure}[hbt!]
	\centering
	\includegraphics[scale=0.55]{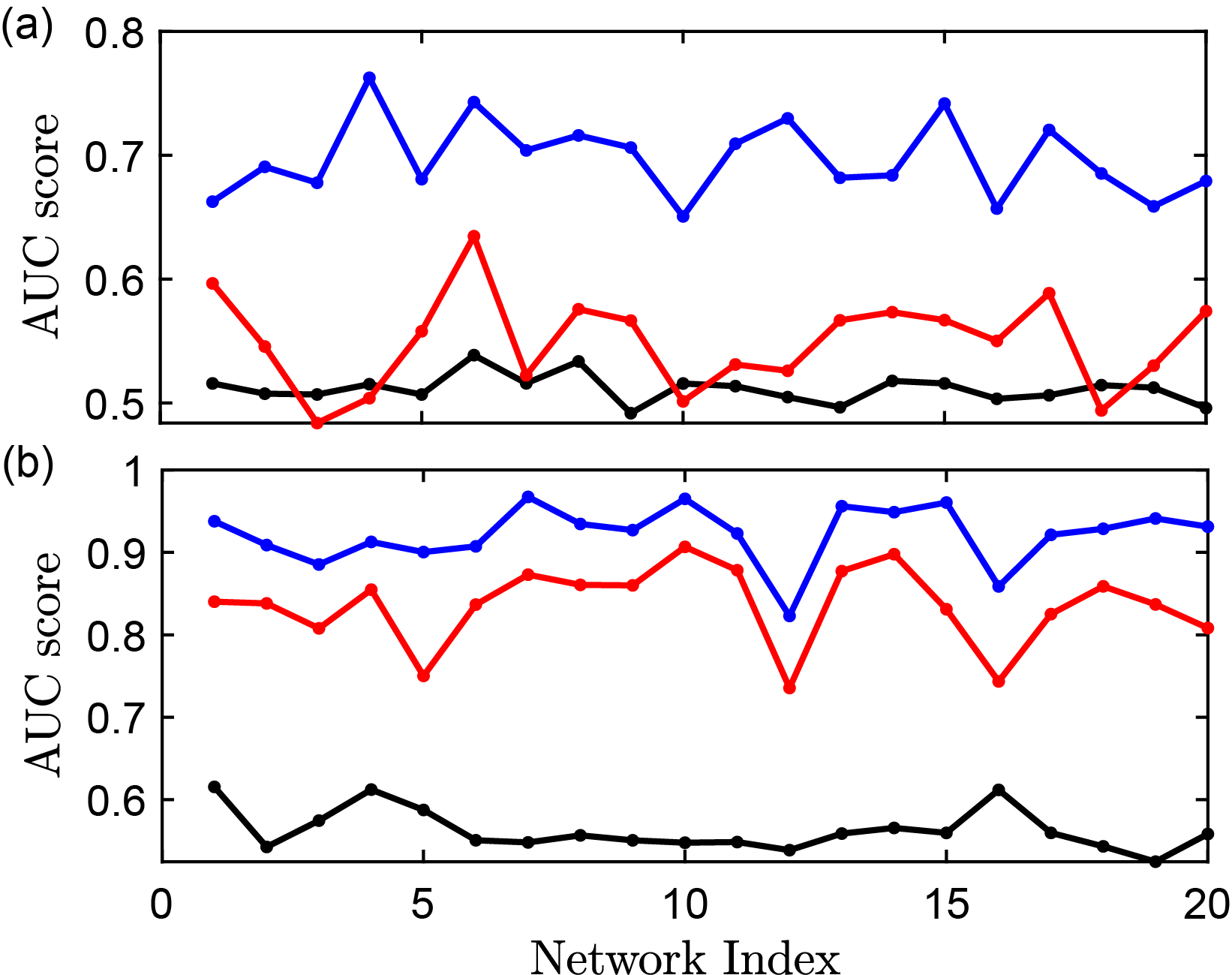}
	\caption{Experimental validation on a network of electronic oscillators. Panels (a) and (b) depict the estimated AUC scores for all 20 networks using the proposed approach (blue), ICON (red), and sparse regression (black) for $k_{ij}=0.01$ and $k_{ij}=0.1$, respectively.}
	\label{fig:figure_5}
\end{figure}

\section{Conclusions}
In this paper, we present a bilinear optimization formulation of the network inference problem by considering the network to have similar coupling functions between the nodes. We then propose an iterative algorithm with sequential initialization with convergence guarantees. The proposed approach is illustrated using the phase model description of oscillator networks; however, the presented ideas can readily be extended to a network of non-oscillatory agents. In addition, the idea of sequential initialization can also be extended to other optimization tasks where solving a system of optimization problems having identical (or similar) solutions is required.

\bibliographystyle{ieeetr}
\bibliography{ref.bib}

\begin{thebibliography}{10}

\bibitem{strogatz2001exploring}
S.~H. Strogatz, ``Exploring complex networks,'' {\em nature}, vol.~410,
  no.~6825, pp.~268--276, 2001.

\bibitem{abel2016functional}
J.~H. Abel, K.~Meeker, D.~Granados-Fuentes, P.~C. St.~John, T.~J. Wang, B.~B.
  Bales, F.~J. Doyle~III, E.~D. Herzog, and L.~R. Petzold, ``Functional network
  inference of the suprachiasmatic nucleus,'' {\em Proceedings of the National
  Academy of Sciences}, vol.~113, no.~16, pp.~4512--4517, 2016.

\bibitem{bomela2020real}
W.~Bomela, S.~Wang, C.-A. Chou, and J.-S. Li, ``Real-time inference and
  detection of disruptive eeg networks for epileptic seizures,'' {\em
  Scientific Reports}, vol.~10, no.~1, p.~8653, 2020.

\bibitem{friston2011functional}
K.~J. Friston, ``Functional and effective connectivity: a review,'' {\em Brain
  connectivity}, vol.~1, no.~1, pp.~13--36, 2011.

\bibitem{lu2021causal}
J.~Lu, B.~Dumitrascu, I.~C. McDowell, B.~Jo, A.~Barrera, L.~K. Hong, S.~M.
  Leichter, T.~E. Reddy, and B.~E. Engelhardt, ``Causal network inference from
  gene transcriptional time-series response to glucocorticoids,'' {\em PLoS
  computational biology}, vol.~17, no.~1, p.~e1008223, 2021.

\bibitem{tibshirani1996regression}
R.~Tibshirani, ``Regression shrinkage and selection via the lasso,'' {\em
  Journal of the Royal Statistical Society Series B: Statistical Methodology},
  vol.~58, no.~1, pp.~267--288, 1996.

\bibitem{wang2018inferring}
S.~Wang, E.~D. Herzog, I.~Z. Kiss, W.~J. Schwartz, G.~Bloch, M.~Sebek,
  D.~Granados-Fuentes, L.~Wang, and J.-S. Li, ``Inferring dynamic topology for
  decoding spatiotemporal structures in complex heterogeneous networks,'' {\em
  Proceedings of the National Academy of Sciences}, vol.~115, no.~37,
  pp.~9300--9305, 2018.

\bibitem{shandilya2011inferring}
S.~G. Shandilya and M.~Timme, ``Inferring network topology from complex
  dynamics,'' {\em New Journal of Physics}, vol.~13, no.~1, p.~013004, 2011.

\bibitem{Singhal2023}
B.~Singhal, M.~Vu, S.~Zeng, and J.-S. Li, ``Data-efficient inference of
  nonlinear oscillator networks,'' in {\em IFAC world congress},
  pp.~10829--10834, 2023.

\bibitem{kiss2002emerging}
I.~Z. Kiss, Y.~Zhai, and J.~L. Hudson, ``Emerging coherence in a population of
  chemical oscillators,'' {\em Science}, vol.~296, no.~5573, pp.~1676--1678,
  2002.

\bibitem{aton2005vasoactive}
S.~J. Aton, C.~S. Colwell, A.~J. Harmar, J.~Waschek, and E.~D. Herzog,
  ``Vasoactive intestinal polypeptide mediates circadian rhythmicity and
  synchrony in mammalian clock neurons,'' {\em Nature neuroscience}, vol.~8,
  no.~4, pp.~476--483, 2005.

\bibitem{schwemmer2012theory}
M.~A. Schwemmer and T.~J. Lewis, ``The theory of weakly coupled oscillators,''
  {\em Phase response curves in neuroscience: theory, experiment, and
  analysis}, pp.~3--31, 2012.

\bibitem{mitrou2017peak}
N.~Mitrou, A.~Laurin, T.~Dick, and J.~Inskip, ``A peak detection method for
  identifying phase in physiological signals,'' {\em Biomedical Signal
  Processing and Control}, vol.~31, pp.~452--462, 2017.

\bibitem{bai2004convergence}
E.-W. Bai and D.~Li, ``Convergence of the iterative hammerstein system
  identification algorithm,'' {\em IEEE Transactions on automatic control},
  vol.~49, no.~11, pp.~1929--1940, 2004.

\bibitem{timme2014revealing}
M.~Timme and J.~Casadiego, ``Revealing networks from dynamics: an
  introduction,'' {\em Journal of Physics A: Mathematical and Theoretical},
  vol.~47, no.~34, p.~343001, 2014.

\bibitem{napoletani2008reconstructing}
D.~Napoletani and T.~D. Sauer, ``Reconstructing the topology of sparsely
  connected dynamical networks,'' {\em Physical Review E}, vol.~77, no.~2,
  p.~026103, 2008.

\bibitem{hanley1982meaning}
J.~A. Hanley and B.~J. McNeil, ``The meaning and use of the area under a
  receiver operating characteristic (roc) curve.,'' {\em Radiology}, vol.~143,
  no.~1, pp.~29--36, 1982.

\bibitem{renyi1960evolution}
P.~E.-A. R{\'e}nyi, ``On the evolution of random graphs,'' {\em Publ. Math.
  Inst. Hung. Acad. Sci. A}, vol.~5, pp.~17--61, 1960.

\bibitem{watts1998collective}
D.~J. Watts and S.~H. Strogatz, ``Collective dynamics of a small-world
  networks,'' {\em nature}, vol.~393, no.~6684, p.~440, 1998.

\bibitem{albert2002statistical}
R.~Albert and A.-L. Barab{\'a}si, ``Statistical mechanics of complex
  networks,'' {\em Rev. Mod. Phys.}, vol.~74, no.~1, p.~47, 2002.

\bibitem{wang2021collective}
J.~Wang and W.~Zou, ``Collective behaviors of mean-field coupled stuart--landau
  limit-cycle oscillators under additional repulsive links,'' {\em Chaos: An
  Interdisciplinary Journal of Nonlinear Science}, vol.~31, no.~7, 2021.

\bibitem{nakao2016phase}
H.~Nakao, ``Phase reduction approach to synchronisation of nonlinear
  oscillators,'' {\em Contemporary Physics}, vol.~57, no.~2, pp.~188--214,
  2016.

\bibitem{vera2020experimental}
V.~Vera-{\'A}vila, R.~Sevilla-Escoboza, A.~Lozano-S{\'a}nchez,
  R.~Rivera-Dur{\'o}n, and J.~M. Buld{\'u}, ``Experimental datasets of networks
  of nonlinear oscillators: Structure and dynamics during the path to
  synchronization,'' {\em Data in brief}, vol.~28, p.~105012, 2020.

\end{thebibliography}

\end{document}